\documentclass[showkeys,superscriptaddress,showpacs,preprintnumbers,aps,twocolumn]{revtex4}

\usepackage{epsfig}
\usepackage{amsfonts}
\usepackage{amssymb}

\newcommand{\ITP}{Institut f{\"u}r Theoretische
Physik, Technische Universit{\"a}t Berlin,
Hardenbergstra{\ss}e 36, D--10623 Berlin, Germany}
\newcommand{\Lancaster}{Department of Physics, Lancaster University,
Lancaster, LA1 4YB, United Kingdom}
\newcommand{\Chemnitz}{Theoretische Physik I, Technische Universit{\"a}t
Chemnitz, D--09107 Chemnitz, Germany}
\newcommand{\QUEENMARY}{School of Mathematical Sciences,
  Queen Mary / Univ.\ of London,
  Mile End Road,
  London E1 4NS, UK}

\begin{document}

\date{\today}

\title{Self-stabilization of high frequency oscillations in
semiconductor superlattices by time-delay autosynchronization}

\author{J. Schlesner}
\affiliation{\ITP}
\author{A. Amann}
\affiliation{\ITP}
\author{N. B. Janson}
\affiliation{\ITP}
\affiliation{\Lancaster}
\author{W. Just}
\altaffiliation{permanent address: \QUEENMARY}
\affiliation{\ITP}
\affiliation{\Chemnitz}
\author{E. Sch{\"o}ll}
\affiliation{\ITP}

\begin{abstract}
We present a novel scheme to stabilize high-frequency domain oscillations
in semiconductor superlattices by a time--delayed feedback loop. 
Applying concepts from chaos control theory 
we propose to control the spatio-temporal dynamics of fronts of accumulation and
depletion layers which are generated at the emitter and
may collide and annihilate during their transit, and thereby suppress chaos.
The proposed method only requires the feedback of internal global electrical 
variables, viz current and voltage, which makes the practical 
implementation very easy.

\end{abstract}


\maketitle

Semiconductor superlattices \cite{ESA70} have been demonstrated to give
rise to self-sustained current oscillations ranging from several hundred
MHz \cite{KAW86,KAS95,HOF96} to 150 GHz at room temperature
\cite{SCH99h}.
Various mechanisms with \cite{BUE77,PAT98,SCH02e,BON02} or without \cite{KRO00} 
the involvement of
propagating field domains have been discusssed. In any case, a superlattice
constitutes a highly nonlinear system \cite{SCH00}, and instabilities
are likely to occur. Indeed, chaotic
scenarios have been found experimentally \cite{ZHA96,LUO98b,BUL99} and described
theoretically
in periodically driven \cite{BUL95} as well as in undriven systems
\cite{AMA02a}.
For a reliable operation of a superlattice as an ultra-high frequency oscillator
such unpredictable and irregular conditions should be avoided.
In principle, synchronization of oscillations in a superlattice by an
external signal \cite{SCH02b} could be exploited to achieve a desired 
periodic behavior. However, 
in reality, the control of the forcing frequency in the ultrahigh range 
presents substantial technical problems.

Here, we propose a simple self-stabilizing scheme that is especially
suitable for semiconductor devices like superlattices. 
It uses a profound concept of chaos control 
from
nonlinear dynamics and chaos theory. Within this approach,  an intrinsically 
unstable time-periodic motion is stabilized using a 
simple feedback loop with a time delay \cite{PYR92}.
This type of control needs only small control forces initially, and they 
vanish once control has been achieved. A sound advantage is that the 
oscillation mode to be stabilized need not be known beforehand,
in contrast to other chaos control schemes.
Rather, a simple delay 
line leads to autosynchronization of the system.
Methods of nonlinear control theory
\cite{OTT90,SCH99c} have been usefully applied to real world problems in
various areas of physics, chemistry and  biology
\cite{BIE94,PIE96,HAL97,SUK97,PAR99,BEN00,LUE01,BEN02,BET03},
but no use has been made of this in the field of semiconductor 
self-oscillators. 

Control methods can be either local or global \cite{BEC02}. Local methods 
require 
our ability to measure, and apply forcing directly to, the spatially 
resolved state variables of the system under 
study. However, in nanotechnology such variables, being e.g. electron 
densities in 
some quantum wells, are not easily accessible, and thus local methods cannot be 
applied. Unlike those, global methods require access only to some 
macroscopic 
variable(s) characterizing some integral output of the system. Such 
output can generally be reliably measured, and 
thus global methods seem to be the only option for the control of devices like 
superlattices. 
However, as we will show below, they are not straightforwardly 
applicable to nano-systems whose structure is spatially discrete. 
In this Letter we present a general approach to self-stabilization of irregular 
oscillations in semiconductor devices based upon essentially 
discrete quantum structures. 

We consider a model for nonlinear electronic transport in semiconductor
superlattices that yields complex and chaotic dynamic behavior
under fixed time-independent external voltage in a regime where 
self-sustained dipole waves \cite{WAC02} are spontaneously generated 
at the emitter. Those dipole waves are associated with traveling field 
domains, and
consist of electron accumulation and depletion fronts that in general
travel at different velocities and may merge and annihilate.
Such moving fronts are widespread in nonlinear, spatially extended systems,
and similar chaotic front patterns occur in many other systems, e.g., 
spatially continuous models describing bulk impurity impact ionization breakdown
in semiconductors \cite{CAN01} or globally coupled heterogeneous catalytic 
reactions \cite{GRA93a}.
Thus the time-delay autosynchronization method
proposed in this work could be readily applied to stabilize similar 
space-time patterns in a variety of systems. 

Our model of a superlattice is based on sequential tunneling of electrons
\cite{WAC02}.
In the framework of this model the quantum wells are assumed to be
only weakly coupled, and electrons are localized at these wells.
The tunneling rate to the next well is lower than the typical
relaxation rate between the different energy levels within one well.
The electrons within one well are then in quasi--equilibrium and
transport through the barrier is incoherent. The resulting
tunneling current density $J_{m\to m+1}(F_m, n_m, n_{m+1})$ from well
$m$ to well $m+1$ depends only on the electric field $F_m$ between
both wells and the electron densities $n_m$ and $n_{m+1}$ in the respective wells
(in units of $cm^{-2}$). 
A typical dependence of $J_{m\to m+1}$ 
on the electric field between two consecutive wells is 
$N$-shaped and exhibits a pronounced 
regime of negative differential conductivity.

The rate of variation of electron density in well $m$ is 
governed by the continuity equation
\begin{equation}
  \label{eq:continuity}
  e\frac{\text{d}n_m}{\text{d}t} = J_{m-1 \to m}  - J_{m \to m+1}
\quad \text{for } m  = 1, \ldots N
\end{equation}
and Gauss's law
\begin{equation}
  \label{eq:poisson}
  \epsilon_r \epsilon_0 (F_{m} - F_{m-1}) = e(n_m -N_D)
\quad \text{for } m  = 1, \ldots N,
\end{equation}
where $N$ is the number of wells in the superlattice, 
$\epsilon_r$ and  $\epsilon_0$ are the relative and absolute 
permittivities, $e<0$ is the electron charge, 
$N_D$ is the donor density, and
$F_0$ and $F_N$ are the fields at the emitter and collector
barrier, respectively.
The total applied voltage $U$ between emitter and collector imposes a
global constraint
$U = - \sum_{m=0}^N F_m d$,
where $d$ is the superlattice period. This, together with 
(\ref{eq:poisson}), allows us to eliminate the field variables
$F_m(n_1, ..., n_N, U)$ from the dynamic equations.

At the contacts Ohmic boundary conditions
$J_{0 \to 1} = \sigma F_0$,
$J_{N \to N + 1} = \sigma F_N n_N/N_D$ are chosen,
where $\sigma$ is the Ohmic contact conductivity, and the factor $n_N/N_D$ is
introduced in order to avoid negative electron densities at the
collector. The value of $\sigma$ essentially determines the 
oscillation mode \cite{AMA02a}. 

\begin{figure}
  \begin{center}
    \epsfig{file=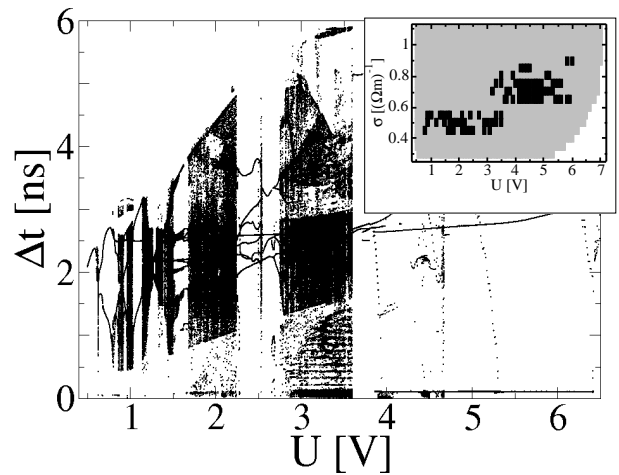,angle=0,width=0.45\textwidth}
    \caption{One-parameter bifurcation diagram:
      Time differences between
      consecutive maxima of the electron density in well no. 20 vs voltage at
      $\sigma=0.5\quad \Omega^{-1}\text{m}^{-1}$. Time series of length
      $600 ns$ have been used for each value of the voltage. The inset shows
      a two-parameter bifurcation  diagram: black squares denote chaotic
      oscillations, light shading indicate periodic oscillations, and
      the white region shows the absence of oscillations.
Simulation of an $N = 100$ superlattice with 
Al$_{0.3}$Ga$_{0.7}$As barriers of width $b=5 \text{nm}$ and GaAs quantum
wells of width $w=8 \text{nm}$, doping density 
$N_D= 1.0 \times 10^{11} \text{cm}^{-2}$ and scattering
induced broadening $\Gamma = 8 \text{meV}$ at $T=20 \text{K}$.  
      }
    \label{fig:positions_delta_t}
  \end{center}
\end{figure}

If the contact conductivity $\sigma$ is chosen appropriately,
electron accumulation and depletion fronts are generated at the emitter.
Those fronts form a traveling high field domain, with leading electron depletion
front and trailing accumulation front. This leads to self-generated current 
oscillations.
A fixed voltage $U$ imposes a constraint
on the lengths of the high-field domains and thus on the
front velocities. If $N_a$ accumulations fronts and $N_d$
depletion fronts are present, the respective front velocities $v_a$ and $v_d$
must obey $v_d/v_a = N_a/N_d$.
If the accumulation and depletion fronts have different velocities, they may
collide in pairs and annihilate.  At certain combinations of contact 
conductivity $\sigma$ and voltage $U$, chaotic motion arises, when the 
annihilation of fronts of opposite polarity occurs at irregular
positions within the superlattice \cite{AMA02a}.  The inset of 
Fig.~\ref{fig:positions_delta_t}
shows the plane of 
$\sigma$ and $U$, where regions with distinct regimes are marked by different 
shading. Black regions are those where chaotic behavior has been found. 
As a computationally convenient criterion for chaos we have 
used the rapid decay of the autocorrelation function estimated 
from $n_{20}(t)$. 
Chaotic regimes
are found at low contact conductivity 
and low voltages, where dipole oscillations
with leading accumulation and trailing depletion fronts occur, and at higher
contact conductivity and higher
voltage, where the role of accumulation and depletion fronts is interchanged. 
In Fig.~\ref{fig:positions_delta_t}, a one-parameter bifurcation diagram 
is given, obtained by plotting the
time differences $\Delta t$ between two consecutive maxima of
the electron density in a specified well. The value of $\sigma$ is 
$0.5\quad \Omega^{-1}\text{m}^{-1}$, and $U$ is 
changed. Chaotic bands and periodic windows can be clearly seen. 

\begin{figure}
  \begin{center}
    \epsfig{file=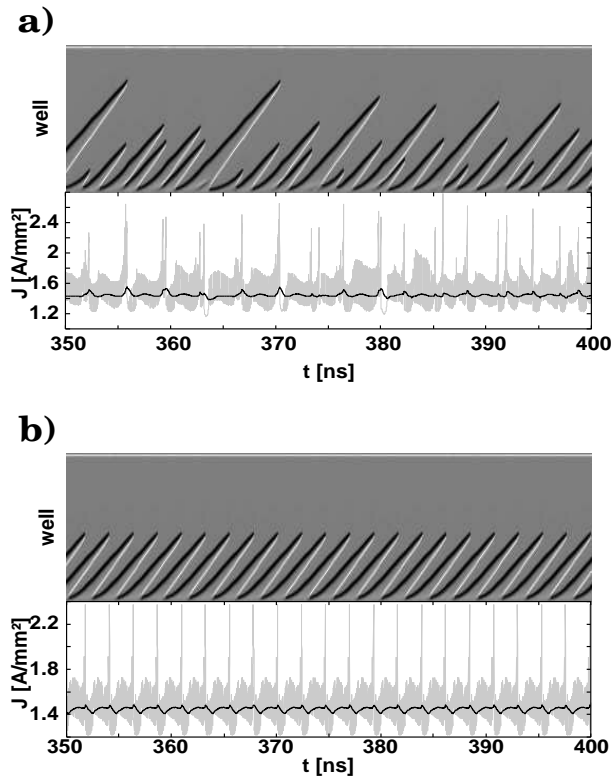,angle=0,width=0.45\textwidth}
\caption{Control of chaotic front dynamics by extended time-delay
autosynchronization. a) Space-time plot of the uncontrolled
charge density, and current density $J$ vs. time.
b) Same with global voltage control with exponentially weighted current
density (denoted by the black curve). Parameters as in Fig.~\ref{fig:positions_delta_t}, $U=1.15$ V,
$\tau=2.29$ ns, $K=3\times 10^{-6}$, $R=0.2$, $\alpha=10^9 s^{-1}$.
      Light and dark regions denote electron
      accumulation and depletion 
      fronts in the space-time plots of the 
      charge densities, respectively.}
   \label{fig2}
  \end{center}
\end{figure}

The transition from periodic to chaotic oscillations is enlightened by
considering the space-time plot for the evolution of the electron densities
(Fig.~\ref{fig2}a).
At $U =1.15 \text{V}$
chaotic front patterns with irregular sequences of annihilation of front
pairs at varying positions within the superlattice occur. We have calculated
the largest Lyapounov exponent as $1.1\times 10^9 s^{-1}$
which is a clear indication of chaos.

We shall now introduce a time delayed feedback loop to control the
chaotic front motion and stabilize a periodic oscillation mode which is
inherent in the chaotic attractor. 
In the extended time--delay autosynchronization scheme 
first suggested by Socolar et al \cite{SOC94},
multiple time delays are used to improve the control performance. 
Analytical insight into those schemes has been gained only recently
\cite{BLE96,JUS97,NAK97}, and various ways of coupling of the control 
force, 
including local and global schemes, have been compared 
\cite{FRA99,BEC02,BAB02,JUS03}. 
Whereas local coupling schemes usually lead to efficient control in a 
large control domain,
they are not easily implemented in real systems since local, spatially 
resolved measurements are necessary. Therefore, here we propose a much
simpler global scheme.
In our problem, as a global output signal that is coupled
back in the feedback loop, it is natural to use the total 
current density $J$ defined 
as follows: $J=\frac{1}{N+1}\sum_{m=0}^N J_{m \to m+1}$ \cite{WAC02}. For 
the uncontrolled chaotic oscillations, $J$ is given in Fig.~\ref{fig2}a by grey, 
showing irregular spikes
at those times when two fronts annihilate. 
Note that the grey current time trace is modulated by fast
small-amplitude oscillations
(due to well-to-well hopping of depletion and accumulation fronts in our
discrete model) which are not resolved in the plot. However, as the variable $J$ is fed back to the system 
for the purposes of control, these high--frequency oscillations render the control loop unstable. 
They need to be filtered out by using e.g. the following low--pass filter:
\begin{eqnarray}
    \overline{J}(t) &=& \alpha \int_0^t J(t') 
e^{-\alpha (t-t')} \mathrm{d}t'
    \mbox{,}
\end{eqnarray}
with a cut-off frequency $\alpha$.

The multiple time delays of the resulting signal $\overline{J}$ (Fig.~\ref{fig2}, black curve) 
are then used to modulate the voltage $U$ across the superlattice:
\begin{figure}
  \begin{center}
    \epsfig{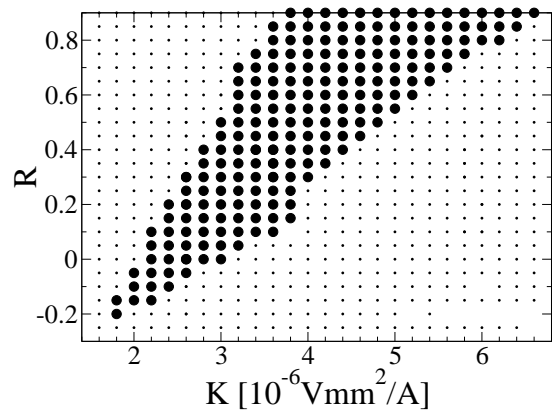}
\caption{Control domain for global voltage control 
with exponentially weighted current density. Full circles
denote successful control, small dots denote no control. Parameters as in 
Fig.~\ref{fig2}.
}
   \label{fig3}
  \end{center}
\end{figure}
\begin{eqnarray}
    \label{global-U}
     U &=& U_0 + U_c(t)  \\ \nonumber
    U_c(t) &=& -  K\left(\overline{J}(t)-\overline{J}(t-\tau)\right) +
    R U_c(t-\tau)\\ \nonumber
    &=& - K \sum_{\nu=0}^{\infty} R^\nu \left( \overline{J}(t-\nu\tau) -
    \overline{J}(t-(\nu+1)\tau)\right) \\ \nonumber
\end{eqnarray}
where $U_0$ is a time--independent external bias, and $U_c$ is the control
voltage. $K$ is the amplitude of the control force, $\tau$ is the delay
time, and $R$ is a memory parameter.
Such a global control scheme is easy to implement experimentally. It is
non-invasive in the sense that the control force vanishes when the
target state of period $\tau$ has been reached. This target state is
an unstable periodic orbit of the uncontrolled system. The period $\tau$ can be
determined by observing the resonance-like behavior of the mean control force
versus $\tau$. The result of the control is shown in Fig.~\ref{fig2}b.
The front dynamics exhibits annihilation of front pairs at fixed positions
within the superlattice,
and stable periodic oscillations of the current are obtained.
In Fig.~\ref{fig3} the control domain is depicted in the parameter plane of $R$ and 
$K$.
A typical horn-like control domain similar to the ones known from other
coupling schemes \cite{BEC02} is found. Typically, the left-hand control 
boundary
corresponds to a period-doubling bifurcation, while the right-hand boundary is 
associated with a Hopf bifurcation. These findings show that our control scheme is 
robust. 

To conclude, we have demonstrated that time-delay autosynchronization 
represents a convenient and simple
scheme for the self-stabilization of high-frequency current oscillations
due to moving domains in superlattices. This approach lacks the drawback 
of synchronization by an external 
ultrahigh-frequency forcing, since it requires nothing but delaying of the 
global electrical system output by the specified time lag. The proposed 
low-pass filtering of the 
output signal presents a solution of the problem one necessarily encounters 
when trying to control a nano-system with a crucially discrete 
quantum structure
leading to superimposed fast well-to-well hopping oscillations in 
our case.

\vspace{0.2cm}
This work was supported by DFG in the framework of Sfb 555 and through
grant no. JU261/3-1. NJ acknowledges partial support by EPSRC.
We gratefully acknowledge discussion with A. Wacker.  

\bibliographystyle{prsty-fullauthor}

\end{document}